\documentclass[12pt]{article}

\usepackage{setspace}

\usepackage{amssymb}
\usepackage{amsfonts, amssymb,amsthm,amsmath, graphics,color,lscape,bm,float,graphicx, subfig}
\usepackage[colorlinks,citecolor=blue,urlcolor=blue]{hyperref}
\usepackage[numbers]{natbib}

\numberwithin{equation}{section}
\theoremstyle{plain}

\newtheorem{thm}{Theorem}[section]
\newtheorem{lemma}{Lemma}[section]

\newtheorem{remark}{Remark}[section]
\newtheorem{note}{Note}[section]

\title{\textbf{Modeling Risk and Return using\\ Dirichlet Process Prior}}

\author{\textbf{Sourish Das\footnote{Sourish Das's research was partially supported by Infosys Foundation grant to CMI. }}$^{a,c}$, \textbf{Aritra Halder}$^b$, \textbf{Ananya Lahiri}$^a$ and \textbf{Dipak K Dey}$^b$ \\
$^a$ Chennai Mathematical Institute, TN, INDIA \\
$^b$ University of Connecticut, Storrs, CT, USA \\
$^c$ University of Southampton, UK
}

\begin{document}
\maketitle

\begin{abstract}
In this paper, we showed that the no-arbitrage condition holds if the market follows the mixture of the geometric Brownian motion (GBM). The mixture of GBM can incorporate heavy-tail behavior of the market. It automatically leads us to model the risk and return of multiple asset portfolios via the nonparametric Bayesian method. We present a Dirichlet Process (DP) prior via an urn-scheme for univariate modeling of the single asset return. This DP prior is presented in the spirit of dependent DP. We extend this approach to introduce a multivariate distribution to model the return on multiple assets via an elliptical copula; which models the marginal distribution using the DP prior. We compare different risk measures such as Value at Risk (VaR) and Conditional VaR (CVaR), also known as expected shortfall (ES) for the stock return data of two datasets. The first dataset contains the return of IBM, Intel and NASDAQ and the second dataset contains the return data of 51 stocks as part of the index ``Nifty 50" for Indian equity markets.
\end{abstract}


\section{Introduction} \label{Intro}

The nature of the log-returns of a financial asset is characterized by heavy-tails, significant skewness and kurtosis. 
Parametric modelling of the data as explained in \cite{Habib2011, 1Shreve2004, 2Shreve2004} concentrate mainly on fitting a normal distribution, generalized hyperbolic distributions, or heavy tail distrubution from extreme value family. This requires us to obtain maximum likelihood estimates (MLEs) for the paramaters of the distribution, which inspite of suffering from overfitting are also inconsistent estimates for the expected return. 
The usual binomial asset pricing model is a discrete time analogue to the continuous time geometric Brownian motion (GBM) \cite{1Shreve2004, 2Shreve2004}. However, empirical studies indicate \cite{HeavyTail2003} that financial log-returns are characterized by heavy-tailed distributions. Using the geometric Brownian motion to model stock/asset prices adversely affects the estimation of quantities like VaR, Conditional Value at Risk (CVaR) and other ``coherent" risk measures \cite{McNeilFrey2000, HeavyTail2003}. A single normal distribution fails to account for the heavy-tails, where a mixture normal ditribution often performs better.  In this paper, we consider mixture of normal distributions as starting point for modeling log-return. Naturally, it motivates us for the mixture of GBM on the stock prices. 

In this context, a nonparametric Bayesian (NPB) \cite{Ferguson1973} approach to modelling has three advantages. The first advantage being data adaptivity. In this paper we present methodology accompanied by an algorithm that takes the data as its only input. Secondly, as we use a DP approach to model the data. The DP essentially fits a finite mixture model with regard to the choice of the base measure. The number of components that are fitted to the data, is learned by augmenting a stochastic process to the algorithm. Finally the tails are accounted for by the finite mixture model that is fitted; since the components of the mixture model vary, the tail behaviour explanation is better explained by changes in the precision parameter of the base-measure.

We develop a multivariate distribution to model the multiple asset's return using a multivariate $t$ copula; which models the marginal distribution using the DP prior. In marginal distribution, the DP prior takes care of heavy-tail behavior of the single asset return and similarly, the multivariate $t$ copula incorporate the heavy-tail behavior of joint distribution of the multiple asset return.

In section \ref{Motivation}, we present the motivation for mixture models. In this section, we showed that no-arbitrage condition holds, if the marke follows mixture of GBM. The mixture of GBM can incorporate heavy-tail behavior. In section \ref{Method}, we aim to establish the framework for the DP prior. We have different subsections, (\ref{SA}) through (\ref{MA-C}); the first deals with the modelling of a single asset using the DP priors. Leter the approach extend to the higher dimensions using an elliptical copula for multiple asset return. In section \ref{RiskMes}, a gradual development to an optimal risk measure is provided indicating the advantages and disadvantages for suggested measures. The performance of these risk measures is assessed based on the suggested probability model for log-returns of assets.  In section \ref{Comp} we elaborate on the computational details that are needed to implement the modelling in both univariate and multivariate cases. Finally, section \ref{App} deals with the application aspects of the suggested modelling approach in univariate as well as multivariate datasets. Section \ref{Conc} concludes the paper.

\section{Motivation for Mixture Models}\label{Motivation}

Suppose $S_t$ is the price of a stock at time point $t$, which follows geometric Brownian motion (GBM). The stochastic differential equation (SDE) corresponding to the GBM is:
\begin{equation*}
    dS_t=S_t(\mu dt+\sigma d B_t), \ S_0=1, 
\end{equation*}
where $B_t$ is Brownian motion, $\mu$ is drift parameter, $\sigma$ is volatility. Then solution of SDE, namely geometric Brownian motion or stock price model, is as follows:
\begin{equation*}
    S_t=S_0 e^{\mu t +\sigma B_t-\frac{1}{2}\sigma^2 t}.
\end{equation*}
The log return is $r_t=\log S_t - \log S_0= \mu t +\sigma B_t-\frac{1}{2}\sigma^2 t$, with expectation 
$\mathbb{E} [r_t]=\mathbb{E} [\log S_t]= \mu t -\frac{1}{2}\sigma^2 t= (\mu -\frac{1}{2}\sigma^2 )t$ and and variance
$\mathbb{V} [r_t]= \mathbb{V} [\log S_t]= \sigma^2 t$. So, we have, $r_t\sim N((\mu -\frac{1}{2}\sigma^2 )t,  \sigma^2 t) $. With this background in mind, we consider the log-return as mixture of Brownian motions, which is as follows:
$$r_t\sim \sum_{i=1}^n \pi_i N((\mu -\frac{1}{2}\sigma_i^2 )t,  \sigma_i^2 t )=\sum_{i=1}^n \pi_i X_i, $$ where 
$\sum_{i=1}^n \pi_i=1$ and $$X_i\sim N((\mu -\frac{1}{2}\sigma_i^2 )t,  \sigma_i^2 t )  $$ with expectation 
$\mathbb{E} [r_t]= \sum_{i=1}^n \pi_i (\mu -\frac{1}{2}\sigma_i^2 )t= (\mu -\frac{1}{2}\sum_{i=1}^n \pi_i \sigma_i^2 )t$ and variance is
$\mathbb{V} [r_t]= \sum_{i=1}^n \pi_i^2\sigma_i^2 t$. Note that, we can express $X_i$ as
\begin{equation*}
  X_i=\mu t +\sigma_i B^i_t-\frac{1}{2}\sigma_i^2 t.
\end{equation*} 
So the log-return can be expressed as, 
\begin{equation*}
    r_t= \sum_{i=1}^n \pi_i X_i =\mu t+ \sum_{i=1}^n \pi_i \sigma_i B^i_t  -\frac{1}{2}\sum_{i=1}^n \pi_i \sigma_i^2 t,
\end{equation*}
where $B^i_t$'s are independent Brownian motions for $i=1,\cdots, n$ on same complete probability space and filtration by $\{B^i_t, \ i=1,\cdots, n\}$ is denoted by $\mathcal{F}_t$.
\begin{thm}\label{thm_martingale}
If $ \tilde{r_t} =r_t-t \big(\mu- \frac{1}{2}\sum_{i=1}^n \pi_i \sigma_i^2\big),$
then $ \mathbb{E} [ \tilde{r}_t |\mathcal{F}_t]= \tilde{r}_{t-1}.$
\end{thm}
\begin{remark}
The $\tilde{r_t} $ is martingale. If $\mu- \frac{1}{2}\sum_{i=1}^n \pi_i \sigma_i^2=r_f $,  where $r_f$ is the risk free interest rate, then $\tilde{r}_t=r_t- t r_f $ can be interpreted as discounted log-return.
\end{remark}

\begin{remark}
This theorem implies if the market follows finite mixture of GBM, then there is no arbitrage opportunity in the market. However, the market is incomplete.
\end{remark}

\section{Methodology for Modelling Asset Return} \label{Method}
The seminal paper by \cite{Sethu1994} provided a constructive definition of the DP-prior and their mathematical properties. Recent advancements in Monte Carlo techniques makes it possible to implement the DP-prior for constructing various kinds of generalized mixture models. The DP is obtained using an infinite-generalization of the finite-Dirichlet distribution. Mixture models consider a kernel, $\mathcal{K}(x,\theta)$ on the space $\mathcal{X}\times \Theta$, where $\mathcal{K}(x,\theta)$ is a measurable function satisfying the condition that for all $\theta \in \Theta$, $\mathcal{K}(\cdot ,\theta)$ is a density with respect to some $\sigma$-finite measure on  $\mathcal{X}$. Let $\Pi$ denote the class of all priors defined on $\Theta$. Then for some $P\in \Pi$, a prior is induced on the density of $\mathcal{K}$, via the map $P \mapsto \mathcal{K}(\cdot ,P)$. The DP priors are useful in providing infinite dimensional extensions to finite dimensional mixture models and consequently assign priors on unknown distributions. The predictive distribution for the problem is then given by,
\begin{eqnarray}\label{bayes-hist}
\theta _1,\ldots,\theta_k &\stackrel{\mathsf{Exch.}}{\sim}& P \in  \Pi,\nonumber\\
X_1,\ldots,X_n |\theta _1,\ldots,\theta_k  &\stackrel{iid}{\sim}& \sum_{i=1}^{k \leq n} h^{-1}K\Big(\frac{x-\theta_i}{h}\Big),
\end{eqnarray}
which essentially approximates $\int K(x,\theta) dP(\theta)$, given $\Big\lbrace P(\theta), X_1, \ldots, X_n \Big\rbrace$. Furthermore, if we can equip $\mathcal{X}$, with natural topologies, like the weak or norm topology, issues related to posterior consistency of estimates obtained using (\ref{bayes-hist}) can be found in \cite{Ghosal1999}. \cite{Zarepour2008} introduced an approach to model Value at Risk (VaR), by assigning DPs priors directly on the log-return of a financial security/asset. According to \cite{Gelman2004}, the DPs are not used directly to model data.  In this paper, we use the mathematical formulation, similar to \cite{Ghosal1999}, with $K$ as the univariate normal distribution and $\Pi$ as the required DP, with the base measure as $\alpha P_0$. This is as an alternative attempt to model financial log-return via a non-parametric Bayesian approach. The assumptions of normality on the base measure $P_0$ does not affect the composition of data, in terms of modelling around the locations.

let us consider $\mathcal{X}=\mathbb{R}$, then we have the corresponding measurable space $(\mathbb{R},\mathcal{B}(\mathbb{R}))$. Let $(\Theta, \mathcal{B}(\Theta))$ denote the corresponding measurable parametric space, then a prior $P \in \Pi$, is a probability measure on $(\Theta, \mathcal{B}(\Theta))$. If $P \sim DP(\alpha P_0)$, be a DP-prior on $(\Theta, \mathcal{B}(\Theta))$ and $B \subset \mathbb{R}$ then,
\begin{eqnarray*}
P(B) | X_1,\ldots,X_n  &\sim& \mathsf{Beta}\Big(\alpha P_0(B)+\sum_{i=1}^{n}\delta_{X_i}(B),\alpha\lbrace 1-P_0(B) \rbrace+\sum_{i=1}^{n}\delta_{X_i}(B^{c})\Big),
\end{eqnarray*}
which holds for all $B\in \mathcal{B}(\mathbb{R})$ \cite{Ferguson1973}. It follows that,
\begin{eqnarray}
E[P(B)| X_1,\ldots,X_n ] &=& \frac{\alpha}{\alpha + n} P_0(B)+ \frac{n}{\alpha + n} \sum_{i=1}^{n}\frac{1}{n}\delta_{X_i}(B),\label{bayes-est}\\
&\to& \sum_{i=1}^{n}\frac{1}{n}\delta_{X_i}(B), ~~ \text{as } n \to \infty \label{bayes-boot1}.
\end{eqnarray}

\subsection{Modelling Single Asset:}\label{SA}

Throughout this paper we shall refer to $S_{t_i}$ as the price-path, $R_{t_i}$ as its corresponding log-return. Associated with $R_{t_i}$ the volatility measure used is the standard deviation of the increment process $\lbrace R_t \rbrace$. Then, $R_{t_i} = \mathrm{ln}\Big(\frac{S_{t_{i}}}{S_{t_{i-1}}}\Big)$, where $S_{t_i}$ is the price of the security at time $t_i$. The model we consider as,
\begin{eqnarray}\label{ret-DP}
R_{t_i} &\sim& \mathcal{K}(\theta),\nonumber\\
\theta &\sim& P,\nonumber\\
P &\sim& DP(\alpha P_0).
\end{eqnarray}
According to \cite{Ferguson1973}, we see that if, $P \sim DP(\alpha P_0)$ is a sample of size 1, then for $B \in \mathcal{B}(\Theta)$, $\Pi(P \in B) =\frac{\alpha P_0(B)}{\alpha P_0(\Theta)}$, therefore,
\begin{eqnarray*}
f_{S_t}(s_t,\theta,P \in B) &=& S_0\cdot e^{\mathcal{K}(R_t|\theta)}\cdot P(\theta) \cdot \frac{\alpha P_0(\theta \in B)}{\alpha P_0(\Theta)}.
\end{eqnarray*}
We refrain from making any assumption on the kernel $\mathcal{K}$, in this section to keep the discussion of the ensuing consequences of (\ref{ret-DP}) as generic as possible. 

\begin{note}
We can consider our base measure to be a Weiner measure, in particular Brownian motion (\cite{2Shreve2004}) when modelling log-returns $R_{t_i}$. Then, the Bayes' risk for our approach is the same as that of the volatiliy parameter, $\sigma$ of the geometric Brownian motion. Given the sample $\lbrace R_{t_1},\ldots,R_{t_n} \rbrace$, DP-prior probability structure allows us to adaptively update the estimate of our base measure. The hierarchy then induces a probability structure which models the underlying stock/asset price. 
\end{note}

Therefore, to explicitly obtain the path of the asset price, we integrate the above over all possible $k$-partions of $\Theta$, and $P$. The stick-breaking construction by \cite{Sethu1994} is then used to provide us with an induced map on $R_{t_i}$
\begin{eqnarray}
\mathcal{P}(R_{t_i}) &=& \sum_{h=1}^{\infty}\pi_h\mathcal{K}(R_{t_i}|\theta_h^{*}),\label{res-ind-map}\\
\theta_h^{*} &\sim& P_0,\label{base-mes}
\end{eqnarray}
where $\pi_h \sim \mathrm{stick}(\alpha)$. \cite{Sethu1994} showed that (\ref{res-ind-map}) is also a DP-prior.

A desirable property of the DP prior on $(\Theta, \mathcal{B}(\Theta))$ is conjugacy, as mentioned in \cite{Ferguson1973} and \cite{Sethu1994}. Let $P$ be a Dirichlet process defined on $(\Theta, \mathcal{B}(\Theta))$ with parameter $\alpha$. Then the conjugacy property states, $\theta_1,\theta_2, \ldots, \theta_n \sim P,~ P \sim DP(\alpha) \Rightarrow P|\theta_1,\ldots, \theta_n \sim DP(\alpha+\sum_{i=1}^{n}\delta_{\theta_i})$. Here $\delta_{\theta_i}$ is the measure assigning probability 1, to $\lbrace\theta |~ \theta \in B \rbrace$. In other words, if $\theta_1,\ldots,\theta_n$, be sampled parameters corresponding to log-returns of the asset, the posterior map given the parameters is also a DP with suitably altered parameters. We now aim to derive the form of the distribution function of the induced probability map on the log-returns $R_{t_i}$. 

\begin{thm}\label{Th1}
For a stochastic process $\lbrace R_t \rbrace$, on the measurable space $(\mathbb{R},\mathcal{B}(\mathbb{R}))$, with parameters, $\boldsymbol \theta \in (\Theta,\mathcal{B}(\Theta))$, if $\Pi$, is a probability measure assigned on $(\Theta,\mathcal{B}(\Theta))$, in particular if $\Pi=DP(\alpha P_0)$, then the distribution function induced on $R_t$ is,
\begin{eqnarray*}
P(R_{t}\in A|\boldsymbol \theta \in B) &=& \frac{\alpha}{\alpha+m} \hat{F}_0^{R_t}(A| \boldsymbol \theta \in  B) +\frac{m}{\alpha+m}\cdot \hat{F}_m^{R_t}(A|\boldsymbol \theta \in  B),
\end{eqnarray*}
where, $\boldsymbol \theta = (\theta_1,\ldots,\theta_m)$ is a realization of size $m$ from $DP(\alpha P_0)$, $A \in \mathcal{B}(\mathbb{R})$, and $B \in \mathcal{B}(\Theta)$
\begin{eqnarray*}
\hat{F}_0^{R_t}(A|\boldsymbol \theta \in  B) &=& \int_{A}  \mathcal{K}(R_{t}|\boldsymbol \theta \in B)dP_0(B),\\
\hat{F}_m^{R_t}(A|\boldsymbol \theta \in B) &=& \hat{F}_m^{R_t}(A|\boldsymbol \theta \in B) = \int_{A}  \mathcal{K}(R_{t}|\boldsymbol \theta \in B) d\Big(\sum_{1}^{m}\frac{1}{m} \delta_{\theta_i}(B)\Big),\\
&=& \sum_{i=1}^{m}\frac{1}{m} \int_{A}  \mathcal{K}(R_{t}|\theta_i) dP_{\theta_i}(B).
\end{eqnarray*}
The last equality is obtained by noting that the integral commutes with the finite $m$-sum.  
\end{thm}

\begin{note}
 Typically, we know that $R_t$ is location invariant. In that context clustering mainly affects the volatility parameter of $\lbrace R_t\rbrace$. Therefore, the clusters may be interpreted as volatility regimes located in the $\lbrace R_t\rbrace$. On the other hand, existence of bull and bear-market trends, affect the location of the process $\lbrace R_t \rbrace$. Altogether a location-scale kernel mixture would therefore do justice to both the mentioned facts in conjunction for modelling the increment process.
\end{note}


The development of the DP-priors is fundamentally based on the Polya urn Processes and Chinese restaurant Processes. In this paper we consider $k \to \infty$, $k$ being the number of urns, and the data points as $n$ balls that are given at the start of the experiment. The $k$-urns in the context of $\mathbb{R}$ can be thought of as a partition. Theoretically the partition size can be infinite. We have a prior $\alpha$ and a base-measure $P_0$, which corresponds to the prior knowledge regarding the urn-occupancy and distribution of occupied urns. We then perform the random experiment of throwing the $n$ balls. Here $\alpha$ serves as the tuning parameter controlling the concentration of balls in urns. $P_0$ serves as probability assigned to the urns/partition. With respect to $P_0$ we should have an idea regarding the furthest expected urn occupancy in our throw. Thus one throw produces $\lbrace 1,2, \ldots, H_0\rbrace \subset \mathbb{N}$ of urns that are occupied. Note that $H_0$ is the maximum number of urn that can be occupied by the $n$ balls.

\begin{lemma}\label{prop1}
For a base-measure $P_0$, a tuning parameter $\alpha$, and the resulting set $\lbrace 1,2, \ldots, H_0\rbrace \subset \mathbb{N}$, for given $H_0 < \infty$, the urn-modelling occurs almost surely in $1,2, \ldots, H_0$ urns across iterations $1,\ldots,T$.
\end{lemma}

\subsection{Modelling Multiple Assets through Copula}\label{MA-C}

Given a collection of $p$ assets, $\boldsymbol S =(S_1,S_2,\ldots,S_p)$, a portfolio (\cite{ArbitMath2006}) is a $p$-vector consisting of the appropriate weights. In this section we assume that we have a given portfolio, that is a set of appropriate weights. We consider the portfolio in terms of the associated log-return for the concerned $p$-assets, $\boldsymbol R =(R_1,R_2,\ldots,R_p)$. We also assume that an investor allocates a fixed sum according to the portfolio $\boldsymbol S$ and the observable prices are at time $\boldsymbol t_n = \lbrace t_1,t_2,\ldots,t_n\rbrace$. The associated log-return over this chosen time horizon, $\boldsymbol t_n$ is denoted by $((R_{it_j}))$, for $i=1,2,\ldots,p$ and $j=1,2,\ldots,n$. We aim to model the marginal distribution of $((R_{it_j}))_{i=1(1)p,~j=1(1)n}$ in this section, using the DP-priors.

The correlations between the $p$-assets can be modelled using the $p$-dimensional multivariate probability distribution. In the previous section (\ref{SA}), we presented the methodology to model the return $R_t$ for a single asset using a DP-prior. In this section we use the copula technique to model the marginal distribution using the  DP-prior. The correlation structure is modelled using the elliptical $t$ copula such as multivariate $t$-copula.

Elliptical copulas correspond to the class of elliptical distributions through the Sklar's theorem. If $F$ denote the multivariate CDF of an elliptical distribution, $F_i$, the marginal of the $i^{th}$-component and its corresponding inverse $F_i^{-1}$ for $i\in 1,\ldots,p$. Then using the Sklar's Theorem the elliptical copula is determined via
\begin{eqnarray*}
C(u_1,\ldots,u_{p}) &=& F\Big[F_1^{-1}(u_1),\ldots,F_{p}^{-1}(u_{p})\Big].
\end{eqnarray*}
The uniqueness of the copula obtained using Sklar's theorem (\cite{Sklar1959}) relies on the assumptions that the marginal(s) $F_i(u_i)$ all have continuous CDFs. This facilitates the application of the probability integral transform on the marginal(s) to formulate the copula. In section (\ref{SA}) we have showed that the CDF induced on $R_t$ is continuous. Consequently, modelling the $p$-assets using a DP prior and using the induced CDFs as marginal(s) ensures the existence of a unique copula. 

The copula is defined given a covariance matrix $\Sigma$. For instance, the Gaussian Copula \cite{JunYan2007} has the following structure,
\begin{eqnarray*}
c(u_1,\ldots,u_p|\Sigma) &=& |\Sigma|^{-\frac{1}{2}} \mathsf{exp}\Big[\frac{1}{2}\mathbf{q}^{T}(I_p-\Sigma^{-1})\mathbf{q}\Big],
\end{eqnarray*}
where $\boldsymbol q = \Big(\Phi(u_1),\ldots,\Phi(u_p)\Big)^{T}$. The measure of association between the $p$-assets being denoted by $\Sigma$. In this paper we consider an appropriate measure of concordance \cite{Scarsini1984} to obtain the entries $((\sigma_{ij}))$, where $i,j=1,\ldots,p$. An interesting property of the family of concordance measures \cite{CopMetfin2004}, is consistency. If $(R_{1n},\ldots,R_{pn})$ is a sequence of continuous random variables with a copula $C_n$, then as the copula converges (pointwise) then the measure of concordance $M_{1,\ldots,p,n}$ also converges. In this paper we use the Kendall's $\tau$ as $M_{1,\ldots,p,n}$ to model the association between $p$-assets under consideration. Considering this with respect to the results in Section (\ref{SA}) we have the following lemma.

\begin{lemma}\label{Lema1}
Let $(R_{1,n},\ldots,R_{p,n}) \sim DP(\alpha P_0)$ and $C_{p,n}$ be the associated copula. Then by uniqueness of the fitted copula $C_{p,n}$ we have,
\begin{eqnarray*}
\lim_{n \to +\infty} C_n(u_1,\ldots,u_p) = C(u_1,\ldots,u_p),
\end{eqnarray*}
then,
\begin{eqnarray*}
M_{R_{1,n},\ldots,R_{p,n}} \stackrel{\mathsf{a.s.}}{\to} M_{R_1,\ldots,R_p},
\end{eqnarray*}
where $\mathsf{a.s.}$ denotes, \textit{almost surely}.
\end{lemma}

\section{Coherent Risk Measures}\label{RiskMes}

In this section we present the coherent risk measures and consider their performance in reference to the log-returns of marginal components of a $p$-asset portfolio being modelled using DP priors. First, we present a discussion regarding the development of coherent risk mesures, followed by a discussion about how the induced probability structure can be incorporated to evaluate portfolios.

Let $\Omega$ be the sample space. Let the map $X : \Omega \mapsto \mathbb{R}$ denote the map corresponding to the loss or gain for a $p$-asset protfolio over the time horizon $[0,T]$. Then this map $X$ is termed as the risk associated with the portfolio. For instance,
\begin{eqnarray}
R_t^{P} &=& \sum_{i=1}^{p} \omega_i R_{it} = \boldsymbol \omega^{T} \boldsymbol R_t, ~~ \sum_{i=1}^p \omega_i= 1.\label{port-ret}
\end{eqnarray}
Then, $R_t^{P}$ can be interpretted as the loss or gain (risk), in terms of return from the $p$-asset portfolio. In general if, $(\Omega, P)$ be a probability space, and $\mathbb{X}$ be the set of all such risk-maps ($\mathbb{R}^{\Omega}$), a risk measure is defined as a fucntion $\rho: \mathbb{X} \mapsto \mathbb{R}$. 

\begin{remark}
According to \cite{Balbas2009}, $\rho(X) > 0$ implies that a positive value is assigned by the measure $\rho$ to the risk $X$. Therefore, $\rho(X)$ is the minimum amount of capital that is to be added to $X$, by investing in the risk-free rate to surpass any level of risk. Conversely, $\rho(X) < 0$ implies that $-\rho(X)$ can be cashed from the current position without any risk.
\end{remark}

Commonly used measures of risk, such as Value at Risk associated with $X$, $VaR_\alpha(X)$ suffers from a variety of deficiencies. These have been identified (\cite{Wang1996}, \cite{Artzner1999}, \cite{Artzner1997}) to formulate a much robust class of measures of risk given by the ``coherent" risk measures. \cite{Artzner1997} states that coherent measures of risk should satisfy 4 properties: (i) sub-additivity (ii) positive-homegeneity (iii) translation invariance and (iv) monotonicity. \cite{Wang1996} changed how portfolio risk was quantified by establishing a generalized theory of coherent measures risk, using distortion functions. Distortion functions are always defined using an associated probability measure with the risk $X$. \cite{Wang1996} also established the Choquet integral expressions \cite{Denneberg1994}, for the commonly used measures of risk. A distortion function is defined with respect to a probability measure $P$ over a measure/probability space $(\Omega,2^{\Omega})$. If $g:[0,1]\to [0,1]$ be an increasing concave function with $g(0)=0$ and $g(1)=1$ then,
\begin{eqnarray*}\label{distort-fun}
\mu &=& g \circ P.
\end{eqnarray*}
The dual distortion function being,
\begin{eqnarray*}
\tilde{g}(u) &=& 1-g(1-u).
\end{eqnarray*}
The risk-measure is defined through the Choquet integral,
\begin{eqnarray}
\rho_g(X) &=& \int X d\mu(X \in A),\nonumber\\
&=& \int_A g \Big[P(X)\Big] dx  - \int_A 1-g \Big[1-P(X)\Big] dx,\nonumber\\
&=& \int_A g \Big[P(X)\Big] dx - \int_{\mathbb{R}\setminus A} 1-g \Big[P(X)\Big] dx,\nonumber\\
&=& E_{\mu(A)}(X). \label{eq1}
\end{eqnarray}
It is evident that the nature of the distortion function affects the ``coherence" of the obtained risk-measure. Also, if we assume that $X$ is $\mu$-integrable, then the distorted risk measure is the expectation under the \emph{re-weighted} probabilities. By using this construction we obtain risk measures that are coherent. 
\begin{remark}
The $VaR_\alpha(X)$ is not a coherent risk measure. The class of coherent distortion risk measures can be further extended to formulating exhaustive distortion risk measures that are both coherent and complete. Completeness (\cite{Balbas2009}), of a distortion risk measure relates primarily to the property of the distortion function $g$ to utilise information from the original loss distribution associated with the risk triplet $(\Omega, P, X)$, where $X$ is the associated risk. 
\end{remark}

Formally, if $X$ be the associated risk variable over $(\Omega, P)$, then $\rho_g(X)$ is a complete distortion risk measure generated by $g$ if,
\begin{eqnarray*}
P(X > x_1) = P(X > x_2) &\Leftrightarrow& \mu\Big((x_1,\infty)\Big) = \mu\Big((x_2,\infty)\Big),
\end{eqnarray*}
where $x_1,x_2 \in [0,\infty)$. In conjunction to this definition, it is important to state two theorems from (\cite{Balbas2009}).
\begin{thm}\label{thrisk1}
For a distorted probability $\mu$ defined by a distortion function $g$, $\rho_g$ is complete is implied, and implied by $g$ is stricly increasing.
\end{thm}
\noindent The proof immediately follows from the definiton of completeness for distorted risk measures.
\begin{thm}\label{thrisk2}
If $\rho_g$ is a distorted risk measure, then it is an exhaustive distortion risk measure if and only if $g$ is concave and strictly increasing. Also it is exhaustive if and only if $g$ is concave and $g(x)< 1$, for all $x <1$.
\end{thm}
\noindent The proof can be found in \cite{Balbas2009}. These theorems establish mainly that for a risk-measure to be complete the distortion fucntion should effectively incorporate all the information in the loss distribution $(\Omega, P, X)$. 

\begin{remark}
: It is clear, that formulation of a risk-measure calls for exercising caution on two fronts viz., the selection of an appropriate loss distribution $P$ to model the risk function and the choice of an appropriate distortion function $g$, to incorporate all of the information in the associated probability $P$ to measure the risk of a position.
\end{remark}

In reference to the methodology developed above for modelling the log-return of a $p$-asset portfolio using DP priors, we now proceed to look at the performance of the aforesaid risk-measures. The DP is a hyperprior with respect to $\boldsymbol R_t$, as it assigns a DP-prior to  $(\Theta,\mathcal{B}(\Theta))$. In the beginning of this section we have shown how the log-return $R_t^P$ is a valid risk (measure of gain or loss in net worth) associated with a portfolio. In the section (\ref{SA}), we have seen that the DP prior induces a probability measure on the log-return $R_t$. Let us consider a collection of $p$ such log-returns corresponding to respective $p$-assets that are modelled using DP priors. Their covariance being accounted for by an appropriate copula as shown in section (\ref{MA-C}). Then we have an induced probability for the probability space, $\Omega=(\mathbb{R}^p,\mathcal{B}(\mathbb{R}^p))$, through a DP-prior $(\Theta^p, \mathcal{B}(\Theta^p)$. The associated risk is given by equation (\ref{port-ret}) on which the DP induces a probability structure. It is more appropriate to consider equation (\ref{port-ret}) given $\boldsymbol \omega$. Moreover, $R_t^P: \mathbb{R}^p \mapsto \mathbb{R}$. Explicitly,
\begin{eqnarray*}
R_t^P &\sim& \text{DP}(\underbrace{\alpha P_0,\ldots, \alpha P_0}_{p \text{ times}}, C),\\
P_0 &\sim& \mathsf{N-Inv-}\chi^2\Big(\mu_0,\frac{\sigma_0^2}{\kappa_0};\nu_0,\sigma_0^2\Big).
\end{eqnarray*}
Here the DP is a multivariate dependent Dirichlet process with marginals as DP-priors, $DP(\alpha P_0)$ and $C$ is the associated copula with an appropriate concordance measure $M$. 


Now we consider the commonly used measures of risk. 
It is important to consider that given a filtration $\mathcal{F}_{M^*}$, upto iteration $M^*$, the associated probability with the loss $R_t$ for a single asset is given by the follwing equation, ($M^*$ being the number of iterations until the mixing RPM estimate is obtatined.)
\begin{eqnarray}
R_t\Big| \mathcal{F}_{M^*} &=& \sum_{h=1}^{{H^{(M)}}^{*}} \pi^{*}_h \mathcal{N}\Big(R_t\Big|\mu_h^{*},{\phi_h^{*}}^{-1}\Big), \label{RPM}
\end{eqnarray}
where ${H^{(M)}}^{*}, \pi^{*}_h, \mu_h^{*}, {\phi_h^{*}}^{-1}$ are Bayes' estimates obtained after succesful convergence for the parameters of the DP prior. 

\subsection{Risk Measures}

\noindent \textbf{Value at Risk: VaR}

\noindent For an appropriate risk $X$, $VaR_{\gamma}(X)$ is defined as,
\begin{eqnarray*}
VaR_{\gamma}(X) &=& \mathsf{sup} \Big\lbrace x \in \mathbb{R} \Big| P(X\geq x) > 1-\gamma. \Big\rbrace
\end{eqnarray*}
It is obtained as a Choquet integral (\cite{Denneberg1994}, \cite{Balbas2009}) by setting,
\begin{eqnarray*}
g(u) &=& \begin{cases}
0 & \text{ if } 0 \leq u < 1-\gamma \\
1 & \text{ if } 1-\alpha \leq u \leq 1. \end{cases}.
\end{eqnarray*}
Using (\ref{eq1}) we have $VaR_\gamma (X)=\int_0^{\infty} g\Big[P(X \geq x)\Big]dx$. Then, $\rho_g(X)=VaR_\gamma (X)$ is simply the $\gamma \times 100$\% quantile of the loss distribution associated with $X$. By assigning the DP prior on the parameter space $(\Theta, \mathcal{B}(\Theta))$, the induced probability distribution on the log-return variable is given by the theorem (\ref{Th1}). The Value at Risk for a single asset portfolio becomes an $\gamma \times 100$\% quantile of the log-return distribution. The problem for estimating $VaR_\gamma (X)$ is then equivalent to estimating the $\gamma \times 100$\% quantile for $P$.
Assuming that the market assumptions for the model hold, we have the GBM model for the log-return for an investment horizon $[0,T]$,
\begin{eqnarray}
R_t &\sim& \mathcal{N}\Big(\Big(\mu-\frac{\sigma^2}{2}\Big)T,\sigma^2 T\Big). \label{BS}
\end{eqnarray}
When comparing the models (\ref{RPM}) and (\ref{BS}) in terms of estimating quantiles we have clear picture regarding the importance of the loss distribution. The $\gamma \times 100$\% quantile for (\ref{BS}) is obtained by solving,
\begin{eqnarray*}
\Phi\Big(\frac{q_\gamma-(\mu-\sigma^2/2)T}{\sigma\sqrt{T}}\Big) &=& \gamma,
\end{eqnarray*}
whereas, for (\ref{RPM}) we solve the following equation for $q_\gamma$, 
\begin{eqnarray*}
\int_0^{q_\gamma}\sum_{h=1}^{{H^{(M)}}^{*}} \pi^{*}_h \mathcal{N}\Big(R_t\Big|\mu_h^{*},{\phi_h^{*}}^{-1}\Big) dR_t &=& \gamma.
\end{eqnarray*}
Here ${H^{(M)}}^{*}$ is finite. Since the sum is finite we have,
\begin{eqnarray*}
\sum_{h=1}^{{H^{(M)}}^{*}} \pi^{*}_h \int_0^{q_\gamma} \mathcal{N}\Big(R_t\Big|\mu_h^{*},{\phi_h^{*}}^{-1}\Big) dR_t &=& \gamma,\\
\sum_{h=1}^{{H^{(M)}}^{*}} \pi^{*}_h \Phi\Big(\frac{q_\gamma-\mu_h^{*}}{\sqrt{{\phi_h^{*}}^{-1}}}\Big) &=& \gamma.
\end{eqnarray*}
$\Phi$, denotes the distribution function for the standard normal. However, there does not exist any closed form expression for the above equation. It is evident that the estimtes for $q_{\gamma}$ will be different in the two cases. The difference being a direct consequence of (\ref{eq1}), which results in a significant change in the estimate of $q_\gamma$. The $VaR_\gamma(X)$ has been known to suffer from numerous deficinecies, the foremost of them being lack of sub-additivity. It is not a convex measure of risk, therefore diversification in terms of assets does not provide room for optimization \cite{Follmer2002}. Despite of these discrepancies it is widely used as a risk measure due to its simplicity in interpretation. 

The reason behind considering quantile estimation for a single asset is to elucidate the significance of the loss distribution $P$. For modelling the risk for a $p$-asset portfolio, we consider univariate modelling of $p$-assets using the equation in (\ref{RPM}) and consider the covariance structure specified by fitting an appropriate copula. The distortion function $g$ for $VaR_\gamma(X)$ remains constant over the interval $[1-\gamma,1]$, which results in $VaR_\gamma(X)$ not being a complete risk measure. This follows from the equivalent condition stated in Theorem (\ref{thrisk1}). Furthermore, the distortion function $g=1$ over the interval $[1-\gamma,1]$ does not make it suitable for being an exhaustive distortion risk measure as well.

\bigskip

\noindent \textbf{ESF/CVAR: Expected Shortfall/Conditional Value at Risk}

\noindent  This measure of risk was first introduced by \cite{Artzner1999}. For an appropriate risk $X$, $CVaR_\gamma(X)$ or $ESF_{\gamma}(X)$ is defined as,
\begin{eqnarray*}
ESF_{\gamma}(X) &=& \frac{1}{1-\gamma} \int_{0}^{1-\gamma} VaR_{u}(X)du,\\
&=& \frac{1}{1-\gamma} \int_{0}^{1-\gamma}\mathsf{sup} \Big\lbrace x \in \mathbb{R} \Big| P(X\geq x) > 1-u \Big\rbrace du,
\end{eqnarray*}
The associated distortion function being given by $g$,
\begin{eqnarray*}
g(u) &=& \begin{cases}
\frac{u}{1-\alpha} & \text{ if } 0 \leq u < 1-\gamma \\
1 & \text{ if } 1-\gamma \leq u \leq 1 \end{cases}.
\end{eqnarray*}
$CVaR_\gamma (X)$ depends on the $VaR_\gamma (X)$ and therefore significant changes are expected, when considering variations in the distribution $P$ from (\ref{BS}) to (\ref{RPM}). The discussion with respect to the improvements is estimation of risk using equation (\ref{RPM}) as the associated loss distribution holds true for $CVaR_\gamma (X)$ as well. In this case $g$ as a distortion function is better, in terms of information content from the loss distribution $P$. Furthermore, it is easy to see that $g$ is \emph{non-decreasing} and concave in nature. Consequently, $CVaR_\gamma (X)$ is a coherent risk-measure.

\bigskip

\noindent \textbf{WT: Wang's Transform}

\noindent Despite of being coherent, $CVaR_\gamma (X)$ suffers from sensitivity towards severity of loss in final net worth, that is higher risk below $\gamma \times 100$\% points of the loss-distribution. This serves as the major downside for $CVaR_\gamma (X)$; moreover, $g$ being non-decreasing ($g=1$ in the interval $[1-\gamma,1]$ ) does not qualify the risk-measure to be a complete one. The Wang's Transform is a valid measure belonging to the class of complete risk-measures. \cite{Wang2000} draws heavily from the general principles establshed in \cite{Wang1996} to suggest a distortion function $g$ that concentrates on symmettric parametric family viz., Normal class of probability measures. 

The advantage of considering a symettric family being reflected in the dual distortion function $\tilde{g}$. The suggested $g$ being given by,
\begin{eqnarray*}
g_r(u) &=& \Phi\Big[\Phi^{-1}(u)+r\Big],
\end{eqnarray*}
where $r$ is the corresponding market price of risk. It follows from the definition of $g_r$, that for $r > 0$ $g_r$ is a concave, and if $r < 0$, $g_r$ is convex. Therefore, with reference to (\ref{thrisk2}) one can easily derive that the distorted risk measure corresponding to $g_r$, with $r>0$ is complete and exhaustive. Coupled with the stated properties, using equation (\ref{RPM}) as an alternative to (\ref{BS}) provides better estimates for the risk-measure associated to $R_t^P$. 


\section{Computational Issues}\label{Comp}

In this section we consider computational aspects for applying the suggested approach to model the data. We use the blocked Gibbs sampler as an MCMC algorithm that is used to update the cluster specific parameters. The advantages of using the blocked Gibbs sampling can be summarized into two factors. Firstly, instead of using just a scale-prior we can now use location-scale families of DPs to model the data. The blocked Gibbs sampler is suited specifically for the purpose of simultaneously updating multiple parameters. Secondly, we have a conjugate prior for $\alpha$ for the blocked Gibbs in general. This makes the application more data-adaptive and generalized in nature. This section is divided into two parts. The first discussion is about the alterations proposed in case of irregular clusters. This will be preceded by a short digression explaining what are regular clusters with respect to the current theoretical setup. The second discussion mainly features an MCMC algorithm to implement the procedure.

\noindent We make the following assumptions,
\begin{eqnarray}
\mathcal{K}(\theta_h^{(m)}) &=& \mathcal{N}(\mu_h^{(m)},{\phi_h^{(m)}}^{-1}), \label{kernel}\\
P_0 &=& \mathsf{Normal-Inv-}\chi^2 \Big(\mu_0,\frac{\sigma^2_0}{\kappa_0};\nu_0,\sigma^2_0\Big). \label{conj-prior}
\end{eqnarray}
In light of the (\ref{conj-prior}), a subtle yet serious issue is the formation of improper clusters. Broadly we are faced with the following cases: (i) $n_h^{(t)}=0$, (ii)  $n_h^{(t)}=1$, and (iii) $n_h^{(t)}>1$. The second case shows the presence of an improper cluster, for which second order moments loose interpretability. 


\begin{remark}
Let $m \in \lbrace 1, \ldots, M\rbrace$, and $n_h^{(m)}$ denote the number of points allocated in cluster $h$ for a particular iteration $m$. If $n_h^{(m)}=1$, we say that for the $m^{\text{th}}$-iteration the cluster-$h$ is irregularly occupied. The cluster-$h$ looses its usual interpretability in terms of moments. 
\end{remark}

Let us assume that, for a particular iteration, $\boldsymbol \theta ^{(m)} = ~\{ \theta_1^{(m)},\ldots,\theta_H^{(m)}\}$, are the unique values of $\theta \in \Theta$. This characterizes the data $(\theta_k^{(m)},R_{t_j})$ where $\theta_k^{(m)} \in \boldsymbol \theta ^{(m)}$ and $j=1,\ldots,n$. Then for $P|\boldsymbol \theta^{(m)}$, such that $P\sim DP(\alpha P_0)$, we have the distribution function from (\ref{Th1}),
\begin{eqnarray*}
\mathbb{P}\Big(\theta\Big|\boldsymbol \theta ^{(m)}\Big)&=&\frac{\alpha}{\alpha+n} P_0(\theta)+ \frac{n}{\alpha+n} \sum_{j=1}^{H \leq n}\frac{1}{n} \delta_{\theta_j^{(m)}}(\theta),\\
&=& \frac{\alpha}{\alpha+n} P_0 (\theta)+\frac{n}{\alpha+n} \sum_{j=1}^{H \leq n}\frac{1}{n} \delta_{\theta_j^{(m)}}(\theta).
\end{eqnarray*}
If $\sum_{j=1}^{H} n_j =n$,
\begin{eqnarray}\label{true-ret-dist}
  \mathbb{P} (R_t|\theta, \boldsymbol \theta ^{(m)}) 
&=& \frac{\alpha}{\alpha+n} \mathcal{K}(R_{t}|\theta, \boldsymbol \theta ^{(m)})P_0(\theta|\boldsymbol \theta ^{(m)})+ \Big( \sum_{j=1}^{H \leq n} \frac{n_j}{\alpha+n} \mathcal{K}(R_{t}|\theta=\theta_j^{(m)})\Big),
\end{eqnarray}
which clearly shows that the data will tend to cluster in $H$ clusters characterized by $\boldsymbol \theta^{(m)}$. Thus, according to \cite{Gelman2004}, fitting such a prior to the parameter space, should favor clustering on the financial log-return data. 

\subsection{The Algorithm}

Here we present the MCMC algorithm that is used to implement the approach presented in the previous sections. The algorithm is presented using the assumptions made in (\ref{kernel}) and (\ref{conj-prior}); the steps are as follows:

\medskip

\noindent \emph{(i)} \textbf{Setting Hyper-parameters}

\begin{enumerate}
\item Select an appropriate $\epsilon>0$; consequently, and $H(\epsilon)$ and initialize $V_h=\frac{1}{H}$.
\item Set hyperprior values for $a_{\alpha},b_{\alpha},\mu_0,\kappa_0,\nu_0,\sigma_0^2$.
\item Set $P_0$ as the Normal-Inverse-$\chi^2$ conjugate prior for the normal kernel (\ref{kernel}).
\end{enumerate}

\noindent \emph{(ii)} \textbf{MCMC Posterior Updates}

\begin{enumerate}
\item For $h \in 1,\ldots,H$ $\pi_h=V_h\prod_{l<h}(1-V_l)$ and update the parameters for the multinomial sampling by,
\begin{eqnarray*}
p_h &=& \frac{\pi_h\mathcal{N}(R_{t_i}|\mu_h,\phi_h^{-1})}{\sum_{h=1}^{H}\pi_h \mathcal{N}(R_{t_i}|\mu_h,\phi_h^{-1})}.
\end{eqnarray*}
Then draw a sample of size $n$ from $\mathsf{Multinom}(p_1,\ldots,p_H)$. Assign $\theta_1,\ldots,\theta_n$ to $R_{t_1},\ldots,R_{t_n}$ to formulate $(\theta_k^{(m)},R_{t_j})$, with $\theta_k^{(m)} \in \boldsymbol \theta ^{(m)}$ and $j=1,\ldots,n$.
\item Update the precision parameter using the conjugacy of the blocked Gibbs,
\begin{eqnarray*}
a^{(m)} &=& a_{\alpha}+H_{0m}^{\mathsf{max}}-1,\\
b^{(m)} &=& b_{\alpha}-\sum_{h=1}^{H_{0m}^{\mathsf{max}}-1}\mathsf{log}(1-V_h),\\
\alpha &\sim& \mathsf{Gamma}(a^{(m)},b^{(m)}).
\end{eqnarray*}
\item Calculate the cluster occupancy using,
\begin{eqnarray*}
n_h^{(m)} &=& \sum_{i=1}^{n}\delta_{\theta_i=h}.
\end{eqnarray*}
\item For $h \in 1,\ldots,H$ update the stick weights $V_h$ from their beta distributions,
\begin{eqnarray*}
V_h &\sim& \mathsf{Beta}\Big(1+n_h^{(m)},\alpha+\sum_{k=h+1}^{H}n_k^{(m)}\Big).
\end{eqnarray*}
\item For $h \in 1,\ldots,H$,
\begin{itemize}
\item $n_h^{(m)}=0$, resample from $P_0$.
\item $n_h^{(m)}=1, t=1$, 
then the posterior update of prior $\mathsf{N-Inv}-\chi^2\Big(\mu_0,\frac{\sigma_0^2}{\kappa_0};\nu_0,\sigma_0^2\Big)$.
\item If $n_h^{(m)}=1, t>1$, 
then the posterior update of prior $\mathsf{N-Inv}-\chi^2\Big(\mu_0,\frac{\sigma_0^2}{\kappa_0^{(m-1)}};\nu_0,\sigma_0^2\Big).$
\item  $n_h^{(m)}>1, t\geq 1$,
then the posterior update for prior is,
$$\mathsf{N-Inv}-\chi^2\Big(\mu_0^{(m-1)},\frac{{\sigma_0^2}^{(m-1)}}{\kappa_0^{(m-1)}};\nu_0^{(m-1)},{\sigma_0^2}^{(m-1)}\Big).$$
\end{itemize}
These are the posterior Gibbs updates using the previous iterations posterior as the prior for the next. Note that $m=0$ is just the starting value for the parameters.
\item Repeat this until $\alpha$ stabilizes to obtain the mixing RPM.
\end{enumerate}

\subsection{Cluster Regularization}

In the previous section we have seen that the estimate(s) of the RPM and related quantities are simply bootstrap estimates. We can estimate the augmented information for $m =1,\ldots,M^*$ when considered for a fixed $h \in \lbrace 1,\ldots, H_0 \rbrace$. For instance, we have
\begin{eqnarray*}
E\Big(n_h^{(M^*)}\Big|\mathcal{F}_{M^*}\Big) &=& \frac{1}{M^*}\sum_{m=1}^{M^*}n_h^{(m)},
\end{eqnarray*}
as the estimate for the expected cluster occupancy given the filtration $\mathcal{F}_{M^*}$. Thus, for a fixed $(h,m)$ if an irregular cluster in located we do not halt the MCMC procedure, since immediate inference from the posterior in terms of interpretability is not required. Therefore, in a collective manner over all $M^*$-iterations the process remains regular. The interpretation makes sense when considering the possibility of a sample of size 1 from a cluster $h$. In particular, this is the case for extreme observations or outliers. This approach allows us to make room for heavy-tails in the RPM. This being indicated by sparsely occupied extreme clusters.

\section{Application} \label{App}

In this section, we present the application of  the proposed methodology to two different datasets. The section is broken into two different parts. Firstly, we consider univariate modeling of an asset using the DP prior. We present the estimates along with their respective confidence intervals. This is done for three different stocks, namely IBM, Intel, and NASDAQ. Secondly, we consider multivariate modeling of a dataset consists of a portfolio consisting of IBM, Intel, and NASDAQ. Following which a multivariate application is carried out on a much larger dataset consisting of an optimized portfolio over $p=51$ assets from the National Stock Exchange of India (NSEI). Note that these 51 stocks make up the index, ``Nifty 50" for Indian stock markets. For optimizing the portfolio, we use a mean-variance optimization \cite{Markowitz1952} to select a suitable portfolio. We use an appropriate fitted elliptical $t$-copula to account for the correlation structure amongst the 51 stocks in the portfolio.


\subsection{Risk and Return Analysis for Single Asset:}\label{uni-model}

\begin{figure}[h!]
\caption{Plot showing DP-fit against Black-Scholes over the observed log-returns  for NASDAQ, Intel and IBM.\label{Nasdaq}}
\subfloat[Intel]{\includegraphics[width=2in,height=2.5in]{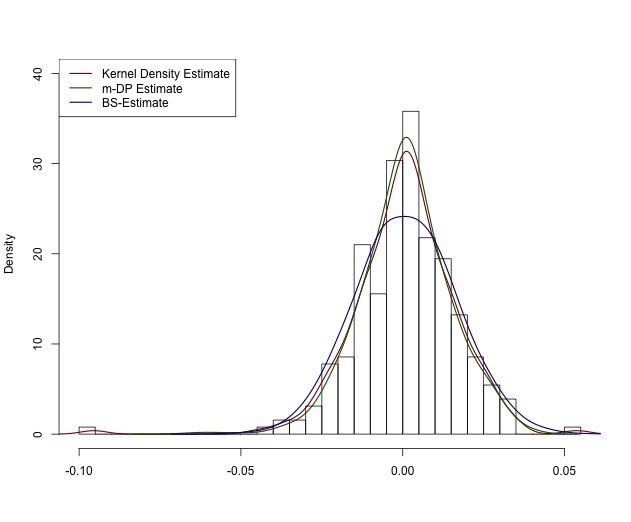}}
\subfloat[IBM]{\includegraphics[width=2.5in,height=2.5in]{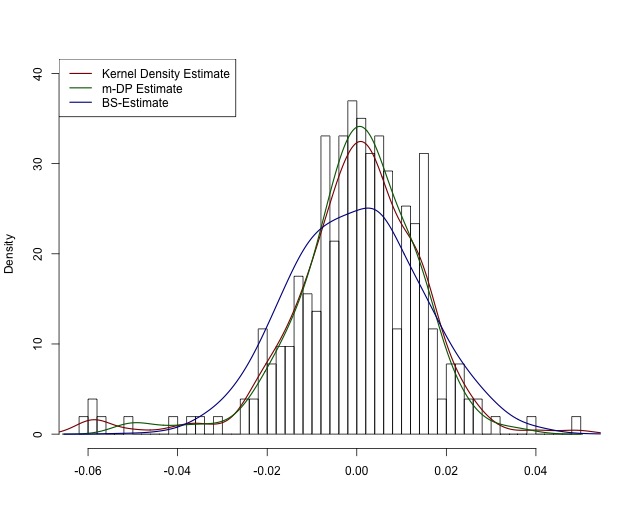}}
\subfloat[NASDAQ]{\includegraphics[width=2.5in,height=2.5in]{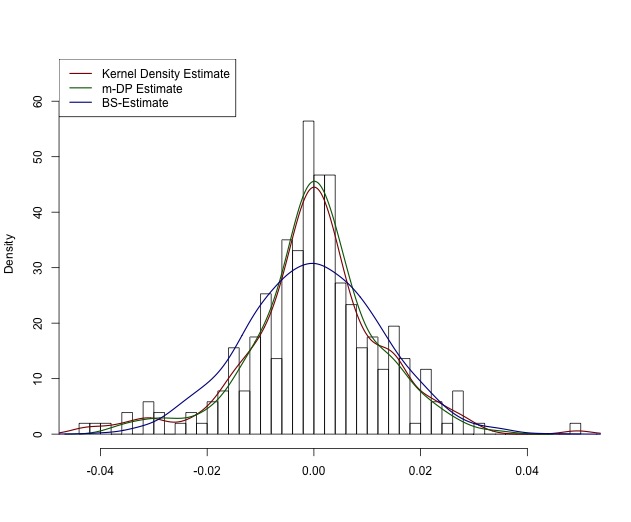}}
\end{figure}

\begin{figure}[h!]
\caption{comparative Fits of Dirichlet Process variations.\label{comphist}}
\subfloat[comparative Fit to IBM]{\includegraphics[width=2.5in,height=2.5in]{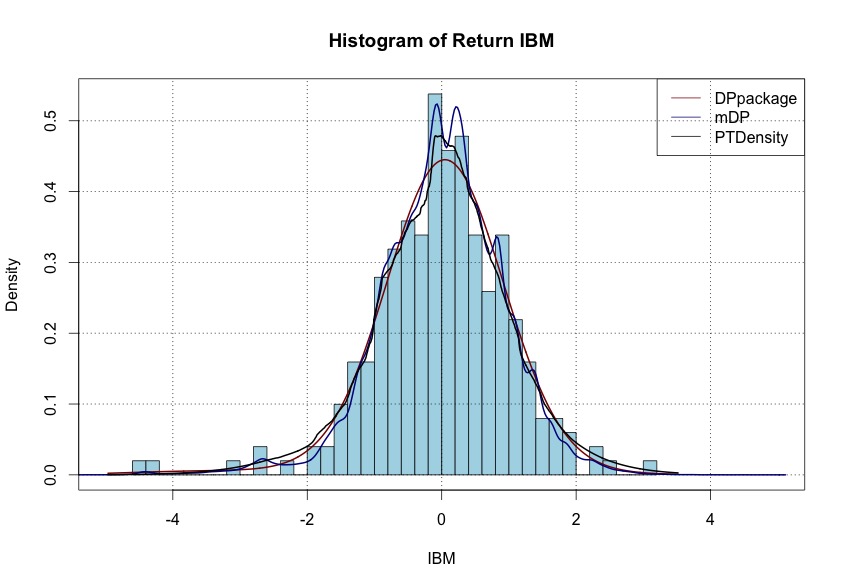}}
\subfloat[comparative Fit to Intel]{\includegraphics[width=2.5in,height=2.5in]{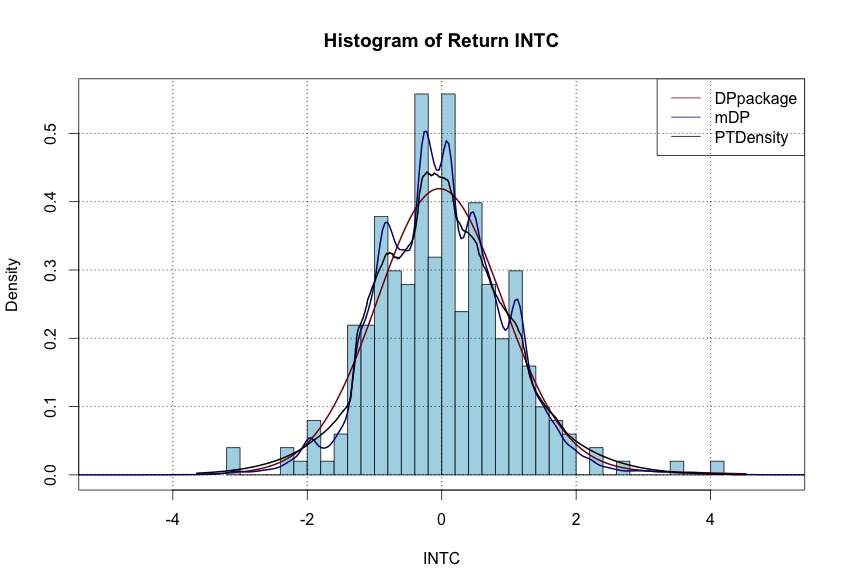}}\\
\subfloat[comparative Fit to NASDAQ]{\includegraphics[width=5in,height=3.5in]{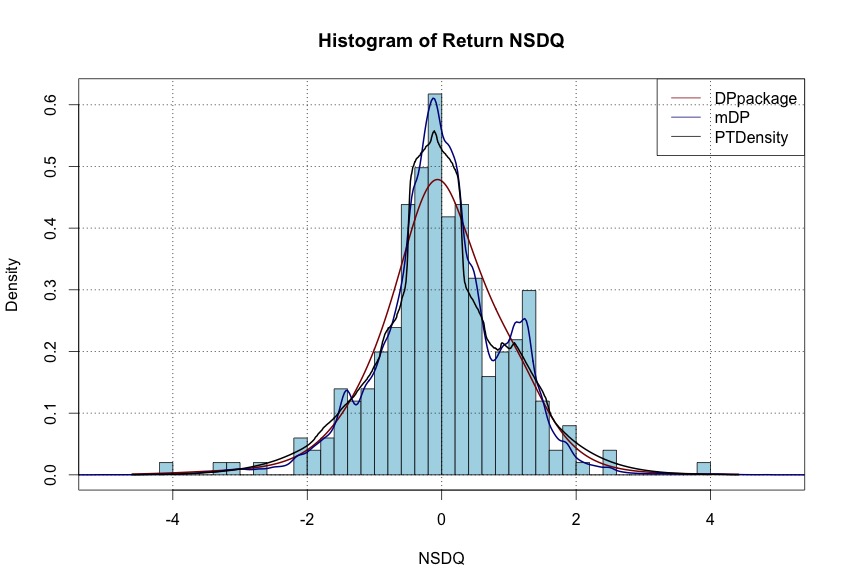}}
\end{figure}

The log-return of the IBM and Intel and the NASDAQ index are modeled using the DP prior and the figure (\ref{Nasdaq}) showing the DP fit against Black-Scholes to Intel, IBM and NASDAQ daily log-returns. The data is collected for these three stocks for a year starting from 1 $^{\text{st}}$ July, 2015. Overall there are 246 days of log-return for the three assets. Simple visual inspection is enough to conclude that DP fit models the log-return much better in comparision to Black-Scholes. Table (\ref{Tbl-comp}) presents comparative capabilities of different methods for density estimation and return path modeling of the three assets. A comparison with the default DP-density and Polya Tail Free priors from \texttt{DPpackage} in \texttt{R} \cite{DPpackage} is shown in Figures (\ref{comphist}). We also calculate the Highest Posterior Density (HPD) intervals in Table (\ref{hpd-est}) with $\alpha=0.1$. It can be seen that the DP prior results in posterior intervals with shortest length with a probability of 1 of contating the mean return and volatility estimate for the underlying asset. In the comparative fits shown in  Figures (\ref{comphist}), we see that the tendency to detect modes and changes in tail behaviour is increased in the m-DP prior. The default \texttt{DPdensity} fails to identify modes completely, while the \texttt{PTdensity} shows modes in the 2-$\sigma$ interval remaining neutral to changes in the tail behaviour. 

\begin{remark}
Considering the kernel density estimate as a benchmark, we compare the performances of the Black-Scholes and DP simulated returns based on mean square deviations. Table (\ref{FigT}) presents a comparative study for the same. Table (\ref{tbl_3_stcks_example}) compares the estimates of the various risk-measures discussed in section \ref{RiskMes}, obtained under the empirical method of modelling log-returns using the kernel density estimate and the multivariate methodology presented.
\end{remark}

\subsection{Risk and Return Analysis for Multiple Asset:}

\begin{figure}[h!]
\centering
\caption{Observed and Simulated return from t-copula for NASDAQ, IBM and Intel\label{scatter_hist}}
\begin{tabular}{cc}
\includegraphics[width=2.5in,height=2.5in]{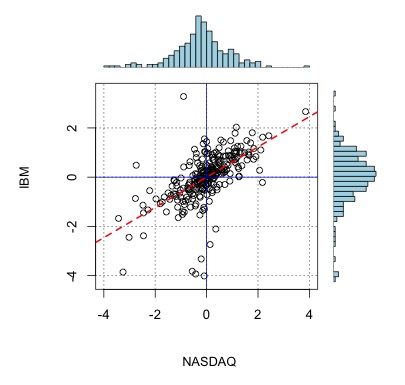}
&
\includegraphics[width=2.5in,height=2.5in]{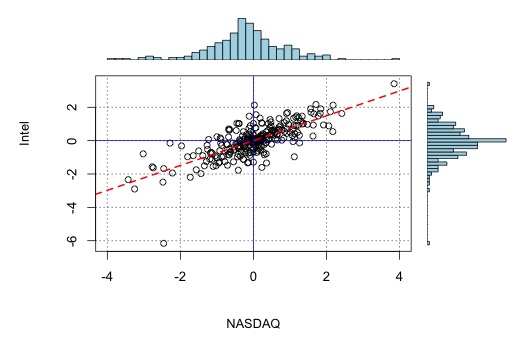}\\
(a) Observed return for NASDAQ vs IBM & (b) Observed return for NASDAQ vs Intel\\
\includegraphics[width=2.5in,height=2.5in]{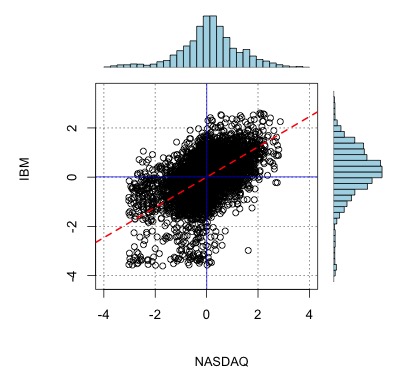}
&
\includegraphics[width=2.5in,height=2.5in]{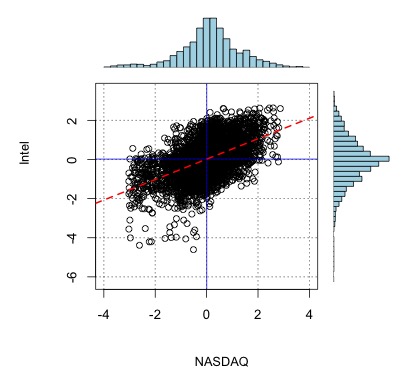}\\
(c) Simulated return for NASDAQ vs IBM  & (d) Simulated return for NASDAQ vs Intel
\end{tabular}
\end{figure}

\begin{figure}[h!]
\centering
\caption{3-D Scatter-plot of Observed and Simulated return from t-copula for Intel, IBM and NASDAQ\label{scatter3}}
\begin{tabular}{cc}
\includegraphics[width=2.5in,height=2.5in]{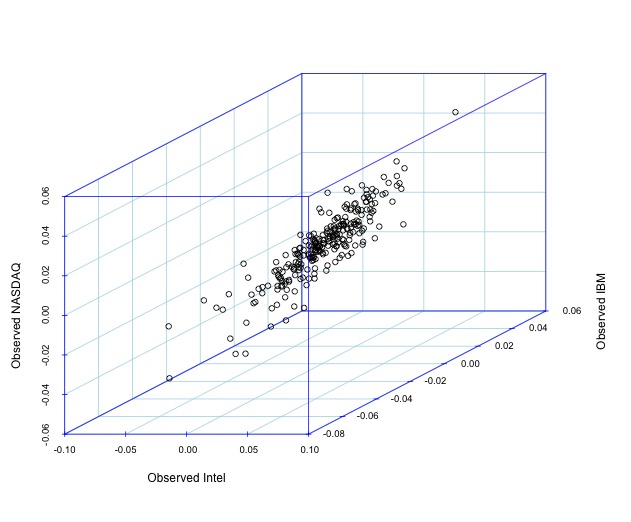}
&
\includegraphics[width=2.5in,height=2.5in]{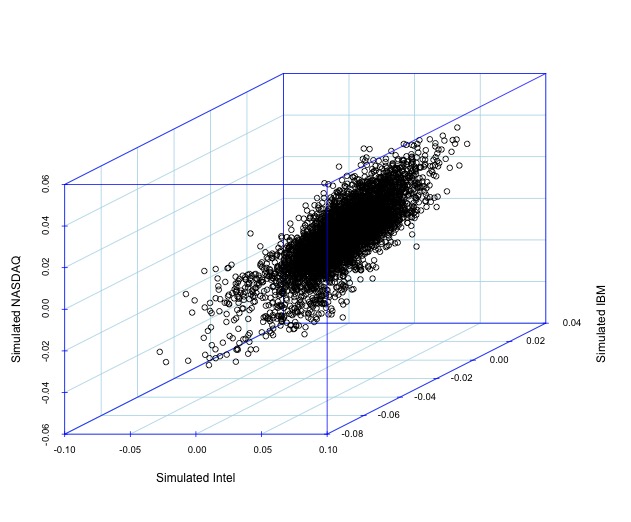}\\
(a) Observed return for Intel, IBM and NASDAQ   & (b) Simulated return for Intel, IBM and NASDAQ
\end{tabular}
\end{figure}

Here we present the application of the multivariate modeling of log-return using the method as discussed in section (\ref{MA-C}). Here we conducted two exercises viz., first we apply the methodology over three assets as considered in (\ref{uni-model}), then we apply the same over the dataset with 51 stocks from Indian stock market. We estimate the covariance matrix $\Sigma$ for $t$ copula using the \cite{Das.Dey.2010, Das.Halder.Dey.2017}. After modelling the three assets using a $t$-copula, we simulate the individual returns with respect to modelled correlation structure. Figures in (\ref{scatter_hist}) show the scatter plots for the observed and fitted daily log-returns for the three assets. The performance of the t-copula, in modelling the observed correlation structure is compared in the figure (\ref{scatter3}). Visual inspection reveals that the observed correlation structure is preserved in simulated log-returns.
 
\begin{figure}[h!]
\centering
\caption{Plot showing PCA on Observed Return and Simulated return\label{PCA}}
\subfloat[PCA-observed]{\includegraphics[width=2.5in,height=2.5in]{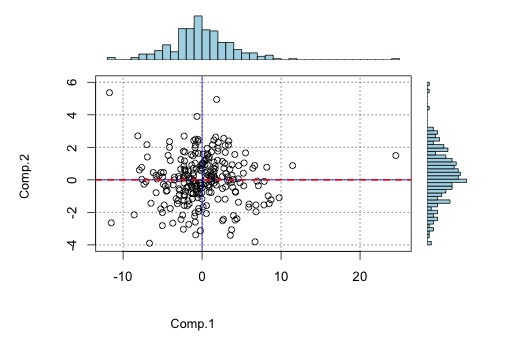}}
\subfloat[PCA-simulated]{\includegraphics[width=2.5in,height=2.5in]{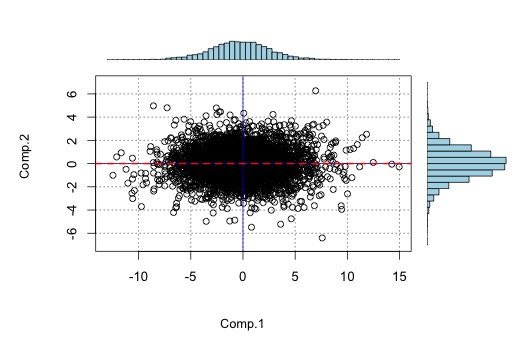}}
\end{figure}

Next, while modelling 51 assets we compare performance of the methodology over the first two principal components for the observed and simulated returns. Figure (\ref{PCA}) shows the scatter plots for the first two principal components in the observed and simulated data respectively. The returns were simulated from a t-copula with 10 degrees of freedom, where the marginals were modeled using the DP prior approach discussed above. Visual inspection indicates that the proposed methodology is able to model the variation in the 51 stocks; along the first two major directions of the correlation structure in the data.

\section{Conclusion} \label{Conc}

In this paper, we presented the Dirichlet Process (DP) prior for modeling the log-return of the single asset(s). In comparison to the approach of \cite{Zarepour2008} we have proposed following alterations. First, we assign a DP prior over parameter space which helps us to avoid inducing a discrete RPM over the log-return. Note that \cite{Zarepour2008} induces a discrete RPM over the log-return \emph{almost surely}. Consequently, we face the problem of quantile estimation for the log-returns, which was avoided by \cite{Zarepour2008}. To deal with this problem, we develop the necessary results that assure the fitting of an RPM, which is \emph{almost-surely a continuous finite mixture} over the log-return; all the while preserving the hierarchy. The proposed results rely heavily on the urn-scheme approach for interpreting DP. For a given set of observations over a fixed time horizon, theoretically assigning a DP involves an infinite mixture modeling. We used the conjugate structure provided by the blocked-Gibbs sampler to augment stochastic processes unique to each question. This augmentation technique helps us to avoid the reversible jump MCMC. 

We extend this approach to introduce a multivariate distribution to model the return on multiple assets via a $t$-copula; which models the marginal using the DP prior. This helps us to keep the already existing nonparametric univariate approach the same even in multivariate applications. The application of this methodology comprises of fitting RPMs over univariate and collection of univariates with a $t$-copula in two different datasets. We compare different risk measures such as Value at Risk (VaR) and Conditional VaR (CVaR) in both the datasets.


\bibliographystyle{authortitle}
\bibliography{bibliography}

\section*{Appendix: Proof}

\noindent \textbf{Proof of Theorem } \ref{thm_martingale}
\begin{proof}
\begin{eqnarray*}
&& E [r_t|\mathcal{F}_{t-1}]\\
&=& E [ \mu t+ \sum_{i=1}^n \pi_i \sigma_i B^i_t  -\frac{1}{2}\sum_{i=1}^n \pi_i \sigma_i^2 t| \mathcal{F}_{t-1}]\\
&=&  E [\mu t+ \sum_{i=1}^n \pi_i \sigma_i (B^i_t-B^i_{t-1})  +\sum_{i=1}^n \pi_i \sigma_i B^i_{t-1} -\frac{1}{2}\sum_{i=1}^n \pi_i \sigma_i^2 t| \mathcal{F}_{t-1}]\\
&=&  \mu t+ \sum_{i=1}^n \pi_i \sigma_i B^i_{t-1} -\frac{1}{2}\sum_{i=1}^n \pi_i \sigma_i^2 t + E [\sum_{i=1}^n \pi_i \sigma_i (B^i_t-B^i_{t-1})| \mathcal{F}_{t-1}]\\
&=&  \mu (t-1)+ \sum_{i=1}^n \pi_i \sigma_i B^i_{t-1} -\frac{1}{2}\sum_{i=1}^n \pi_i \sigma_i^2 (t-1)+ (\mu- \frac{1}{2}\sum_{i=1}^n \pi_i \sigma_i^2)\\
&=& r_{t-1}+(\mu- \frac{1}{2}\sum_{i=1}^n \pi_i \sigma_i^2)
\end{eqnarray*}
So, we get \begin{equation}
E [ r_t-t(\mu- \frac{1}{2}\sum_{i=1}^n \pi_i \sigma_i^2)| \mathcal{F}_{t-1}]=
r_{t-1}-(t-1)(\mu- \frac{1}{2}\sum_{i=1}^n \pi_i \sigma_i^2)
\end{equation}
\end{proof}

\noindent \textbf{Proof of Theorem } \ref{Th1}

\begin{proof}
We consider the $L^{2}$-norm to keep the results analytically tractable. If $B \subset \Theta$, $P_{\theta}\sim DP(\alpha P_0)$ and $F(B) = P_{\theta}(\theta \in B)$ then we have, $L(P,\hat{F}) = \int \Big(F(B)-\hat{F}(B)\Big)^2dW(B),$
$W$ being a finite measure on $(\Theta, \mathcal{B}(\Theta))$. This implies,
$
\mathbb{E}(L(P,\hat{F})) = \int \mathbb{E}\Big(F(B)-\hat{F}(B)\Big)^2dW(B),
$
which is minimized if $\hat{F}(B) = \mathbb{E}[F(B)]$. But, $\mathbb{E}[F(B)]= \mathbb{E}[P_{\theta} \in B]$. From (\ref{bayes-est}) we have,
$\mathbb{E}[F(B)] = P_0(B) \Rightarrow ~ \hat{F}_0(B) = P_0(B),$
which is the prior guess at the shape-scale-location structure of the parameter space. If we have a sample of size $m$, $\boldsymbol \theta =~ (\theta_1,\theta_2, \ldots, \theta_m)$ from $P_{\boldsymbol \theta}$ then by the Bayes' rule,
\begin{eqnarray}\label{DP-DF}
\hat{F}(B|\boldsymbol \theta) &=& w_m \hat{F}_0(B) + (1-w_m) \hat{F}_m(B|\boldsymbol \theta),
\end{eqnarray}
where $w_m=\frac{\alpha}{\alpha+m}$ and
\begin{eqnarray}\label{bayes-boot2}
\hat{F}_m(B|\boldsymbol \theta)= \sum_{1}^{m}\frac{1}{m} \delta_{\theta_i}(B),
\end{eqnarray}
which is also termed as the Bayesian bootstrap estimate for the distribution function \cite{Gelman2004}. The induced map on $R_t$, we have for $A \subset \mathbb{R}$ and $B \subset \Theta$,
\begin{eqnarray*}
P(R_{t}\in A |\boldsymbol \theta \in B) &=& \int_{A} \mathcal{K}(R_{t}|\boldsymbol \theta) dF_\theta(\boldsymbol \theta \in B),\\
&=& \int_{A}  \mathcal{K}(R_{t}|\boldsymbol \theta) d\Big[w_m \hat{F}_0(B) + (1-w_m) \hat{F}_n(B|\boldsymbol \theta)\Big],\\
&=& w_m \cdot \int_{A}  \mathcal{K}(R_{t}|\boldsymbol \theta)d\hat{F}_0(B)\\
&& + ~ (1-w_m) \cdot \int_{A}  \mathcal{K}(R_{t}|\boldsymbol \theta) d\hat{F}_m(B|\boldsymbol \theta),\\
&=& w_m \cdot \int_{A}  \mathcal{K}(R_{t}|\boldsymbol \theta)dP_0(B)\\
& & + ~ (1-w_m) \cdot \int_{A}  \mathcal{K}(R_{t}|\boldsymbol \theta) d\Big(\sum_{1}^{m}\frac{1}{m} \delta_{\theta_i}(B)\Big),\end{eqnarray*}
where $w_m = \frac{\alpha}{\alpha+m}$. If we denote $\hat{F}_0^{R_t}(A|\boldsymbol \theta \in  B)=\int_{A}  \mathcal{K}(R_{t}|\boldsymbol \theta \in  B)dP_0(B)$, and $\hat{F}_m^{R_t}(A|\boldsymbol \theta \in  B)=\int_{A}  \mathcal{K}(R_{t}|\boldsymbol \theta \in  B) d\Big(\sum_{1}^{m}\frac{1}{m} \delta_{\theta_i}(B)\Big)$, then we have,
\begin{eqnarray}
P(R_{t}\in A|\boldsymbol \theta \in  B) = \frac{\alpha}{\alpha+m} \hat{F}_0^{R_t}(A | \boldsymbol \theta \in  B) +\frac{m}{\alpha+m}\cdot \hat{F}_m^{R_t}(A|\boldsymbol \theta\in  B).\label{ind-DP-ret}
\end{eqnarray}
Explicitly, the equation above is the map induced by the DP prior on $\boldsymbol \theta$, through the hierarchy which is referred to in (\ref{ret-DP}). In particular equation (\ref{ind-DP-ret}) is the distribution function of the induced Dirichlet map.

The induced distribution in (\ref{ind-DP-ret}) holds for a given subset $\boldsymbol \theta \in B \subset \Theta$, and $R_t \in A \subset \mathbb{R}$. To generalize the above analysis over the $\sigma$-field $\mathcal{B}(\Theta) \times \mathcal{B}(\mathbb{R})$, we use the $\pi-\lambda$ theorem for a fixed $A \subset \mathbb{R}$ to show that the (\ref{ind-DP-ret}) holds for all $B \in \mathcal{B}(\Theta)$ and vice-versa. 
\end{proof}

\noindent \textbf{Proof of Lemma} \ref{prop1}

\begin{proof}
We initially consider an infinite partition of $\mathbb{R}$ for modelling the $n$ data points. With respect to the base-measure $P_0$ and tuning parameter $\alpha$.
Note that here some $n_h^{(1)}=0$, for $h=1,\ldots,H_0$. The importance of the prior $\alpha P_0$ in selecting such an $H_0$ is to ensure that the sets of form,
\begin{eqnarray*}
\mathcal{L}_{H_0+k}&=&\lbrace h | n_h^{(1)}>0, h = H_0+k+1,\ldots,\infty \rbrace,
\end{eqnarray*}
have a sequence of probabilities under our prior probability information $\alpha P_0$ which tends to be infinitesimal at $H_0 \to k \to \infty$. From the above formulation we also have $\mathcal{L}_{H_0} \supseteq \mathcal{L}_{H_0+1} \supseteq  \mathcal{L}_{H_0+2} \supseteq \ldots$, therefore for all $m \in \lbrace 1,\ldots, M^*\rbrace$. 
\begin{eqnarray*}
P(H>H_0^{(m)}|\boldsymbol P_m) & \stackrel{\mathsf{a.s.}}{\to} & 0.
\end{eqnarray*}
When looked across iterations $1,\ldots,M^*$ we utilize the product measure and the almost sure convergence with respect to $\boldsymbol P_1 \times \ldots  \times \boldsymbol P_{M^*}$ to establish the proposition. Hence, if there exists a set $\mathcal{L} \subset \otimes_{m=1}^{M^*} \mathcal{F}_m$, where the $n$ data points, repeatedly ``binned" across $M^*$ iterations into $H_0\times M^*$ bins lie outside, for some $m, h, n_h^{(m)}$, then the set $\mathcal{L}$ is a set of measure 0, with respect to the product measure $\boldsymbol P_1 \times \ldots  \times \boldsymbol P_{M^*}$ above.  
\end{proof}

\noindent \textbf{Proof of Lemma} \ref{Lema1}

\begin{proof}
The result in lemma (\ref{prop1}) shows that there exists $H_{0i}$, for $i=1,\ldots,p$ such that acorss $1,\ldots,M^*$ iterations the marginal CDFs of $R_{1,n},\ldots,R_{p,n}$ are given by the result in (\ref{Th1}). The CDFs obtained are continuous. If be the copulas $C_{p,n}(\mathsf{max} H_{0i})$ then by Sklar's theorem we have that $\exists$ unique $C$ such that,
\begin{eqnarray*}
\lim_{n \to +\infty}C_{p,n}(\mathsf{max} ~ H_{0i}) = \lim_{n \to +\infty} C_n(u_1,\ldots,u_p,\mathsf{max}~(H_{0i})) &=& C(u_1,\ldots,u_p),
\end{eqnarray*}
almost surely-$\mathsf{max}~(H_{0i})$. Let, $M_{R_{1,n},\ldots,R_{p,n}}$ be the associated concordance measure, then by the inherrent consistency property, we have
\begin{eqnarray*}
M_{R_{1,n},\ldots,R_{p,n}} &\stackrel{\mathsf{a.s.}}{\longrightarrow}& M_{R_1,\ldots,R_p}.
\end{eqnarray*}
\end{proof}

\section*{Figures and Tables} \label{FigT}

\begin{table}[h!]
\centering
\caption{Mean Square Deviation for m-DP and Black-Scholes Estimate} 
\label{msdtab}
\begin{tabular}{|l|rrr|}
  \hline
   & Intel & IBM & NASDAQ \\ 
  \hline
  m-DP & 0.043 & 0.037 & 0.034 \\ 
  BS & 0.364 & 0.326 & 0.298 \\ 
   \hline
\end{tabular}
\end{table}

\begin{table}[h!]
\centering
\caption{Comparison between Kernel Density Estimate, Black-Scholes and m-DP}\label{Tbl-comp}
\begin{tabular}{|l|c c|}\hline
Method & Density Estimation & Return path modeling \\ \hline
Kernel Density Estimate & Yes & No \\ 
Black-Scholes & Poorly & Poorly \\
m-DP & Yes & Yes \\ \hline
\end{tabular}
\end{table}

\begin{table}[h!]
\centering
\caption{Empirical and Copula Estimated VaR (1\%) and Expected Shortfall (1\%) (ESF)} 
\label{tbl_3_stcks_example}
\begin{tabular}{|l|rrrr|}
  \hline
    & Intel & IBM & Portfolio & NASDAQ \\ 
  \hline
  Empirical VaR (1\%) & -2.40 & -3.84 & -2.27 & -2.88 \\ 
  Copula Estimated VaR (1\%) & -2.26 & -3.20 & -2.21 & -2.63 \\ 
  Empirical ESF (1\%)  & -3.85 & -3.94 & -3.21 & -3.23 \\ 
  Copula Estimated ESF (1\%)  & -3.01 & -3.39 & -2.61 & -2.82 \\ 
   \hline
\end{tabular}
\end{table}

\begin{table}[h!]
\centering
\caption{HPD-intervals for the expected return and volatility} 
\label{hpd-est}
\begin{tabular}{|c|rr|rr|rr|}
  \hline
  Statistics & Lower 90\% & Upper 90\% & Lower 90\% & Upper 90\%  & Lower 90\%  & Upper 90\% \\ 
  & (mDP) & (mDP) & (PTF) & (PTF) & (DP) & (DP)\\
  \hline
  $\mu$ (IBM) & -0.031 & 0.0392 & -1.1214 & 0.3980 & -1.7245 & 1.3511 \\ 
  $\sigma$ (IBM) & 0.849 & 0.9664 & 1.0765 & 1.4531 & 0.1436 & 0.8473 \\ 
  \hline
 $\mu$ (INTC) & -0.0349 & 0.0289 & -0.3976 & 0.5564 & -1.3955 & 1.6548 \\ 
 $\sigma$ (INTC)  & 0.8780 & 0.9774 & 0.9661 & 1.4567 & 0.1486 & 0.9553 \\ 
 \hline
 $\mu$ (NSDQ) & -0.0350 & 0.0326 & -0.6707 & 0.3402 & -1.3999 & 1.4284 \\ 
 $\sigma$ (NSDQ) & 0.8664 & 0.9729 & 1.0731 & 1.3655 & 0.1623 & 1.3998 \\ 
  
   \hline
\end{tabular}
\end{table}

\end{document}